\documentclass[11pt, conference]{IEEEtran}
\IEEEoverridecommandlockouts
\usepackage{cite}
\usepackage{amsmath,amssymb,amsfonts}
\usepackage{algorithmic}
\usepackage{graphicx}
\usepackage{textcomp}
\usepackage{xcolor}
\usepackage{comment}
\usepackage{subcaption}
\usepackage{soul}
\usepackage{float}
\usepackage{svg}
\def\BibTeX{{\rm B\kern-.05em{\sc i\kern-.025em b}\kern-.08em
    T\kern-.1667em\lower.7ex\hbox{E}\kern-.125emX}}

\usepackage{hyperref}
\usepackage{url}

\hypersetup{
    colorlinks=true,
    linkcolor=blue, 
    urlcolor=blue
    }
    
\begin{document}

\title{A Lean Transformer Model for \\ Dynamic Malware Analysis and Detection}

\author{
\IEEEauthorblockN{
Tony Quertier\IEEEauthorrefmark{1},
Benjamin Marais\IEEEauthorrefmark{1},
Grégoire Barrué\IEEEauthorrefmark{1}
Stéphane Morucci\IEEEauthorrefmark{1},
Sévan Azé\IEEEauthorrefmark{1},
Sébastien Salladin\IEEEauthorrefmark{2}}
\vspace{0.2cm}
\IEEEauthorblockA{\IEEEauthorrefmark{1}Orange Innovation,Rennes, France}

\IEEEauthorblockA{\IEEEauthorrefmark{2}Orange DSEC, Rennes, France}

\IEEEauthorblockA{\footnotesize{mails : firstname.name@orange.com}}}

\maketitle

\begin{abstract}

Malware is a fast-growing threat to the modern computing world and existing lines of defense are not efficient enough to address this issue. This is mainly due to the fact that many prevention solutions rely on signature-based detection methods that can easily be circumvented by hackers. Therefore, there is a recurrent need for behavior-based analysis where a suspicious file is ran in a secured environment and its traces are collected to reports for analysis. Previous works have shown some success leveraging Neural Networks and API calls sequences extracted from these execution reports. 

Recently, Large Language Models and Generative AI have demonstrated impressive capabilities mainly in Natural Language Processing tasks and promising applications in the cybersecurity field for both attackers and defenders.



In this paper, we design an Encoder-Only model, based on the Transformers architecture, to detect malicious files, digesting their API call sequences collected by an execution emulation solution. 
We are also limiting the size of the model architecture and the number of its parameters since it is often considered that Large Language Models may be overkill for specific tasks such as the one we are dealing with hereafter. In addition to achieving decent detection results, this approach has the advantage of reducing our carbon footprint by limiting training and inference times and facilitating technical operations with less hardware requirements. 
We also carry out some analysis of our results and highlight the limits and possible improvements when using Transformers to analyze malicious files.




\end{abstract}

\vspace{0.2cm}

\section*{Introduction}

The emergence of generative AI \cite{minaee2024large} and transformers \cite{vaswani2023attention} opens up a lot of new opportunities in the field of cybersecurity \cite{motlagh2024large}, for both attackers and defenders. Numerous papers have highlighted various ways in which this new technology can be used to automate attacks, such as the creation of malware \cite{pa2023attacker} or phishing websites \cite{hazell2023spear} for instance, but also to improve Red Teams tooling \cite{deng2023pentestgpt} or accelerate mitigations.

In the context of dynamic malware analysis, where a malicious file is executed in a secure environment and information reports are gathered to identify potential misbehaving activities, researchers focus on AI to deal with the problems of detection and classification \cite{Gaber_review_malware}. The question of how to deal with the information contained in these reports is crucial, and several works \cite{app14031015,jindal2019neurlux,Zhang_Qi_Wang_2020,chen2022cruparamer} have identified the importance of API (Application Programming Interface) calls and their arguments.

Recently, Trizna \textit{et al.} leveraged ``Speakeasy" \cite{Speakeasy}, a Windows emulation tool that makes it possible to collect dynamic traces during the analysis of malware and benign Windows PE files with minimal temporal and computational costs \cite{trizna2022quo}. These traces (i.e. reports) are then processed to train machine learning models for malicious behavior detection. Trizna publicly released two of these models: ``Quo Vadis" \cite{trizna2022quo} and ``Nebula" \cite{trizna2023nebula}. 

In this article, we are building upon Trizna works with a focus on a Small Language Model (SLM) to limit our carbon footprint and to reduce training times and computing resources \cite{sml-capgemini}. Even if larger Language Models tend to perform better \cite{kaplan2020scaling}, it has also been showed that a particular attention needs to be devoted to the ratio between the size of the models and the size of the training data \cite{hoffmann2022training}, especially when dealing with very specific tasks.

Our model aims at identifying malware based on API names and their arguments extracted from Speakeasy emulation reports and tokenized in a custom and optimized way. We compare our results with Nebula \cite{trizna2023nebula}, and also train our model on other datasets to investigate our SLM generalization capabilities.

Note that our definition of an Language Model may differ from others since from our point of view, a Language Model encompasses Encoder-Only, Encoder-Decoder and Decoder-Only architectures.

The paper is organized as follows. In Section \ref{sec:methodology} we detail our datasets, the different preprocessing steps and the architecture of our SLM, Encoder-Only model. Section \ref{sec:results} presents several results, while we detail some limitations in Section \ref{sec:limitations}. Finally, Section \ref{sec:future work} gives some insights into future works in order to improve our model.

\section{Methodology}
\label{sec:methodology}

Motivated by the fact that LLMs require significant computing resources, one side objective is to implement a Small Language Model. For specific tasks, it has been shown that it is not always optimal to implement and train a Large Model with a massive amount of data. Leaner Models can match Large Models performance when they are trained with high quality data \cite{gunasekar2023textbooksneed,li2023textbooksneediiphi15}. Focusing on low demanding models has several benefits both from a carbon footprint perspective and from a usability point of view. Since our models are expected to be operated by Orange technical teams, low-demanding energy and resources models are easier and faster to deploy to production.

\begin{figure}[!ht]
    \centering
    \includegraphics[width=0.2\textwidth]{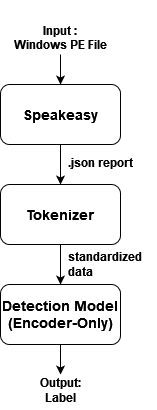}
    \caption{Overview of our detection chain.}
    \label{fig:framework}
\end{figure}

We build an Encoder-Only model, based on the architecture defined in \cite{vaswani2023attention}. We combine Speakeasy emulation reports provided by Quo Vadis and JSON reports generated from two academic datasets and also from our own proprietary corpus. API calls are extracted from these reports, normalized and tokenized using a custom algorithm. Associated tokens are then input into the model for training. 
This model outputs Portable Executable (PE) files \cite{pietrek2002depth} prediction labels based on a maliciousness score. Figure \ref{fig:framework} provides an overview of this detection chain.

\subsection*{Datasets}
\label{subsec:dataset}

To investigate the relevance of a Transformer-based machine learning model achieving malware detection digesting emulation reports, we leverage several datasets in this work. The first one is the freely available Quo Vadis dataset \cite{trizna2022quo} that has been generated using Speakeasy emulation tool and some manual labeling by a professional threat intelligent team. This dataset contains JSON reports of malware and benign execution emulations, with information on API calls, file usages, network traffic and registry accesses, among others. In this work, we are focusing only on API names and their arguments. This dataset has also the advantage of being already split into training and validation corpuses to foster reproducible results. Our model is thus trained on this Quo Vadis train corpus and compared to previous academic results \cite{trizna2022quo,trizna2023nebula}. We also experiment our methodology on the Bodmas dataset provided by \cite{Yang}, with extra benign files extracted from PEMachineLearning dataset \cite{PEML}. This combined dataset is referred to as ``B\&PEML" which stands for ``Bodmas and PEMachineLearning". It finally contains 57,293 malicious files and 81,322 benign files in raw PE format, collected between August 2019 and September 2020. As it is a well-balanced dataset and widely used in the literature \cite{rayankula2023evaluation,hai2023proposed,lu2022self}, it makes sense to evaluate our model on JSON reports generated by Speakeasy on this specific dataset. Comparisons with existing and future works can also be made to benchmark performance. Finally, we use a third internal dataset composed of benign and malicious PE files collected internally during the first months of 2024. This last source of data will give us insights into our model generalization capabilities when facing recent PE files. In this document, we refer to this internal dataset as the OBMID-24 dataset where this acronym stands for ``Orange Benign and Malware Internal Dataset 2024".

Several samples from these datasets produced unusable reports because of too few reported API calls. This is due to the fact that Speakeasy stops its process when it encounters an API call that is not present in its emulation library. In order to prevent some bias related to these early stops, we discard all the reports under 4 KBytes, because they generally contain fewer than five instructions. 
Table \ref{tab:datasets} reports the approximate number of samples per dataset after this filtering.

\begin{table}[!ht]
    \centering
    \caption{Approximate number of retained samples per dataset.}
    \label{tab:datasets}
    \begin{tabular}{|c|c|c|}
        \hline
          & benign & malware \\
         \hline
         Quo Vadis & 33k & 59k \\
         \hline
         B\&PEML & 42k & 39k \\
         \hline
         OBMID-24 & 10k & 5k \\ \hline
    \end{tabular}
\end{table}

\subsection*{Cleaning and normalization}


To facilitate reports processing by our model, we first perform API names and arguments cleaning and normalization. Similarly to Trizna \textit{et al.}. \cite{trizna2023nebula}, we filter out some unnecessary symbols and replace some fields by generic placeholders in order to reduce variability, tokens list size and irrelevant data. For instance, all url addresses are replaced by the $\langle \text{url}\rangle$ placeholder. Table \ref{tab:tags} summarizes several placeholders used during this cleaning and normalization process. This preprocessing limits the creation of pointless tokens, since a tokenizer could for example split an url address into multiple useless parts that may hinder model training and inference efficiencies. 

\begin{table}[!ht]
    \caption{List of some placeholders used to normalize our data.}
    \label{tab:tags}
    \centering
    \begin{tabular}{|c|c|}
        \hline
        Arguments & Tags  \\
        \hline
         http://... & $\langle \text{url}\rangle$ \\
         \hline
         C:\textbackslash\textbackslash... & $\langle \text{path}\rangle$ \\
         \hline
         Long string (100+) & $\langle\text{string}\rangle$ \\
         \hline
         Google\textbackslash\textbackslash Temp & $\langle\text{google}\rangle$ \\
         \hline
         Error 0x80004005 & $\langle\text{error}\rangle$ \\
         \hline 
    \end{tabular}
\end{table}

\subsection*{Tokenization}

Tokenization is a crucial step and must not be overlooked. Even if their work focuses on source code, Jimenez \textit{et al.} \cite{tokenizer-impact} report that differences between tokenizers are of practical importance since the use of different tokenizers may lead to contradictory conclusions. Put another way, a careful selection of tokenizers is required for a given specific task. 

Since API names from Speakeasy reports are constant and finite in number, it does not make a lot of sense to tokenize these names in order to capture different meanings according different contexts. Authors in \cite{effectiveness-ptm-api-learning} found that preventing the tokenizer from decomposing API names can help improve Pre-Trained Models performance. Adding API names as specific tokens to their dictionary makes their model focus more on API sequences rather than on API names alone.  

VMware found similar results in \cite{VMWare-Tokenization} when using common tokenizers algorithms for API names and/or API calls arguments. A WordPiece tokenizer applied to VMware product names, technical jargon and multi-compound words leads to a ``sub-token soup" that hinders Natural Language Processing efficiency.

In the Quo Vadis dataset, we identified about 2,500 unique API names that could all be added to our dictionary making it as large as dictionaries used in other academic works. For instance, Nebula \cite{trizna2023nebula} leverages a 50k tokens dictionary while Neurlux \cite{jindal2019neurlux} relies on a 10k tokens dictionary. However, due to our focus on a limited resources model, we decided to associate specific tokens to top API names only to limit the size of our dictionary. 

We complement this static mapping with different tokens computed by a WordPiece tokenizer \cite{Wordpiece, WPhuggingface} on remaining API names, their associated libraries and all API arguments. WordPiece tokenizer is an efficient solution to alleviate some out-of-vocabulary issues when dealing with rare API names or unseen API parameters. We also double check manually that our dictionary does not exhibit sub-optimal tokenization on all API names and libraries. Our final dictionary size is close to 5,500 tokens.

\begin{figure}[!ht]
     \centering
     \begin{subfigure}[b]{0.5\textwidth}
         \centering
         \includegraphics[width=1\textwidth]{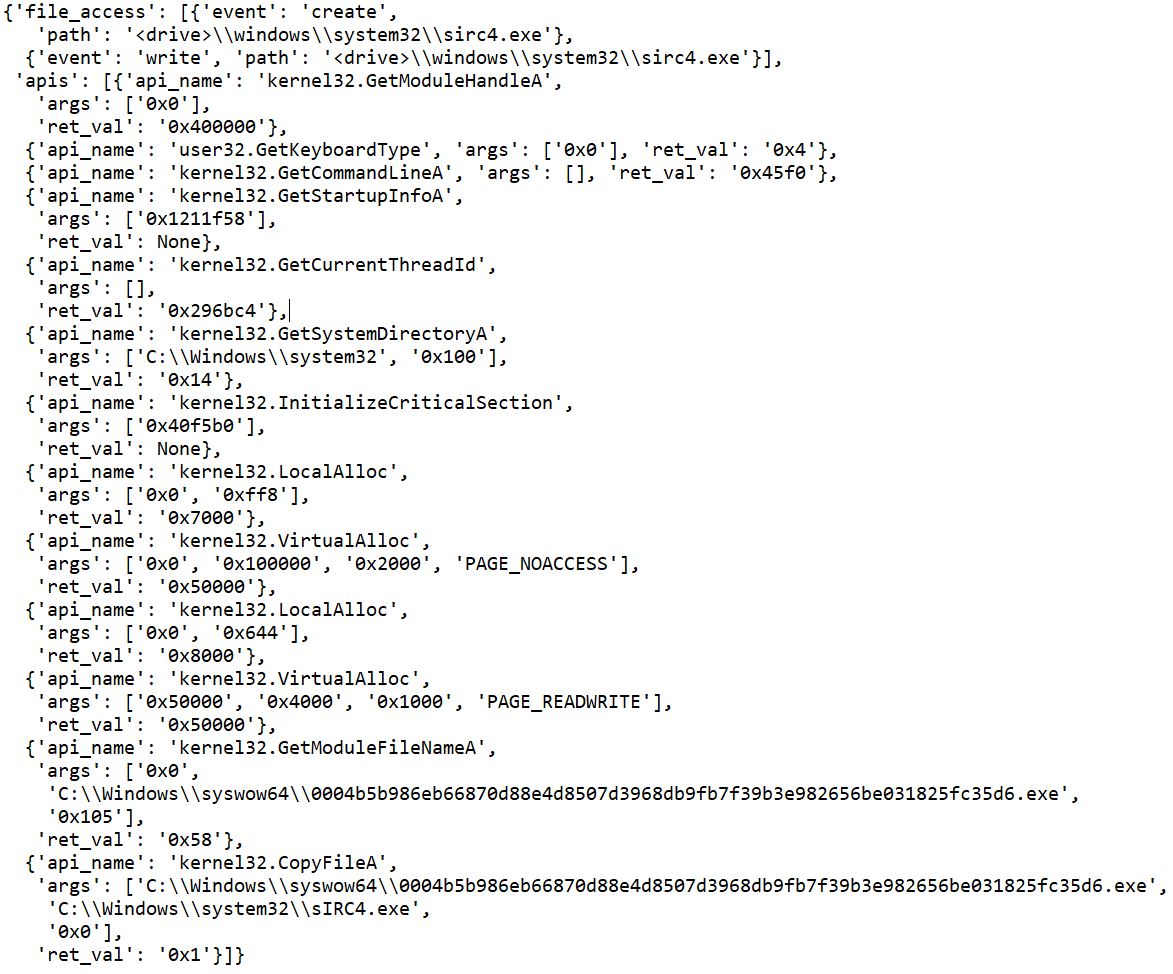}
         \vspace{0.05cm}
         \caption{Original Speakeasy report.}
         \label{fig:preco_report}
     \end{subfigure}
     \hfill     \vspace{0.05cm}
     \begin{subfigure}[b]{0.5\textwidth}
         \centering
         \includegraphics[width=1\textwidth]{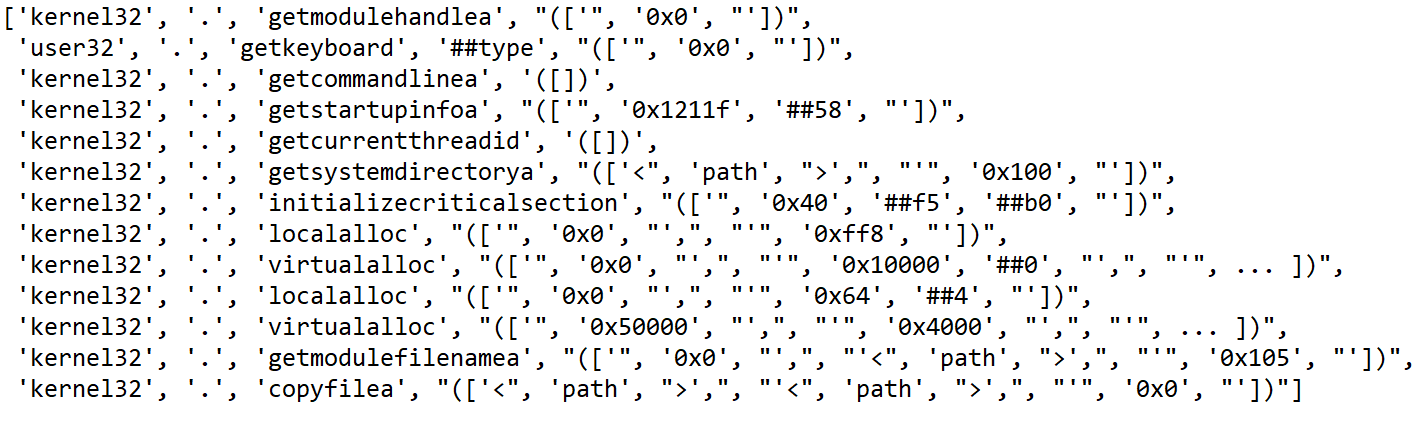}
         \vspace{0.05cm}
         \caption{List of tokens after preprocessing.}
         \label{fig:preco_tokens}
     \end{subfigure}
        \caption{An example of preprocessing applied to a malicious file.}
        \label{fig:preco}
\end{figure}

We present an example of our preprocessing method applied to a malicious file in Figure \ref{fig:preco}. 

\subsection*{Encoder}

We build an Encoder-Only model to classify PE files, based on Speakeasy emulation reports. We decided not to rely on fine-tuning some existing pre-trained Encoder-Only models, like BERT  \cite{devlin2018bert} and its derivatives, for two main reasons: (i) API names and arguments have very different characteristics from the language that BERT model or alike has processed during its training steps and (ii) we want to focus on a relatively modest machine-learning architecture to limit our carbon footprint by reducing time and resources consumption.

\begin{figure}[!ht]
    \centering
    \includegraphics[width=0.30\textwidth]{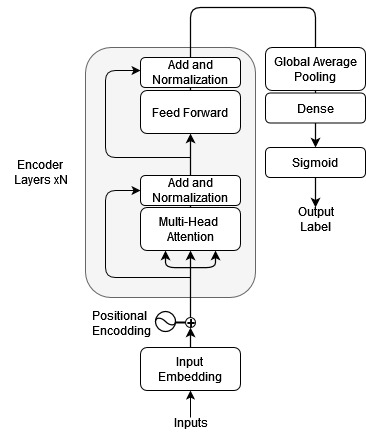}
    \caption{Architecture of our Encoder-Only model.}
    \label{fig:encoder_only}
\end{figure}

Our model is very similar to the Transformer architecture as detailed in \cite{vaswani2023attention}. We are using an embedding layer and some positional encoding before the encoder layers. The encoder layers are composed of multi-head attention (MHA) sublayers, followed by feed-forward neural networks (FFNNs). After each sublayer, we add residual learning \cite{he2015deep} as well as a normalization step to ensure a better learning and a better context understanding. Encoder layers output is then aggregated using a global average pooling and then passed through a fully-connected neural layer and returns the output prediction of the label. Figure \ref{fig:encoder_only} details the different layers of our Encoder-Only model.  

\begin{figure*}[htb]
    \centering
    \begin{subfigure}{0.4\textwidth}
    \includegraphics[width=\textwidth]{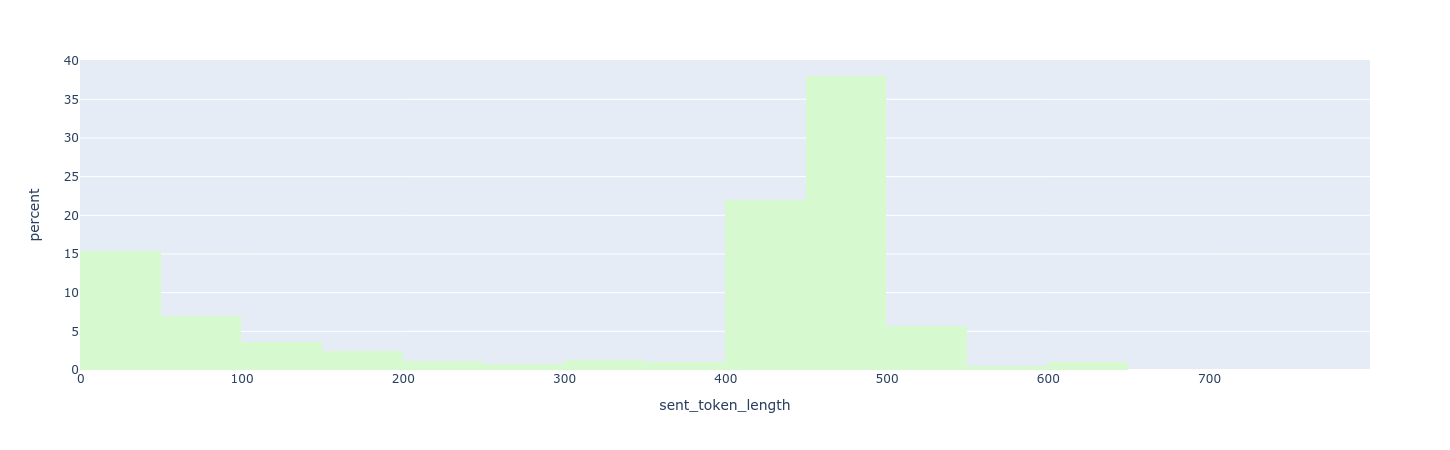}
    \caption{Number of tokens per sequence on Quo Vadis dataset.}    
    \end{subfigure}
    \hspace{0.1cm}
    \begin{subfigure}{0.4\textwidth}
    \includegraphics[width=\textwidth]{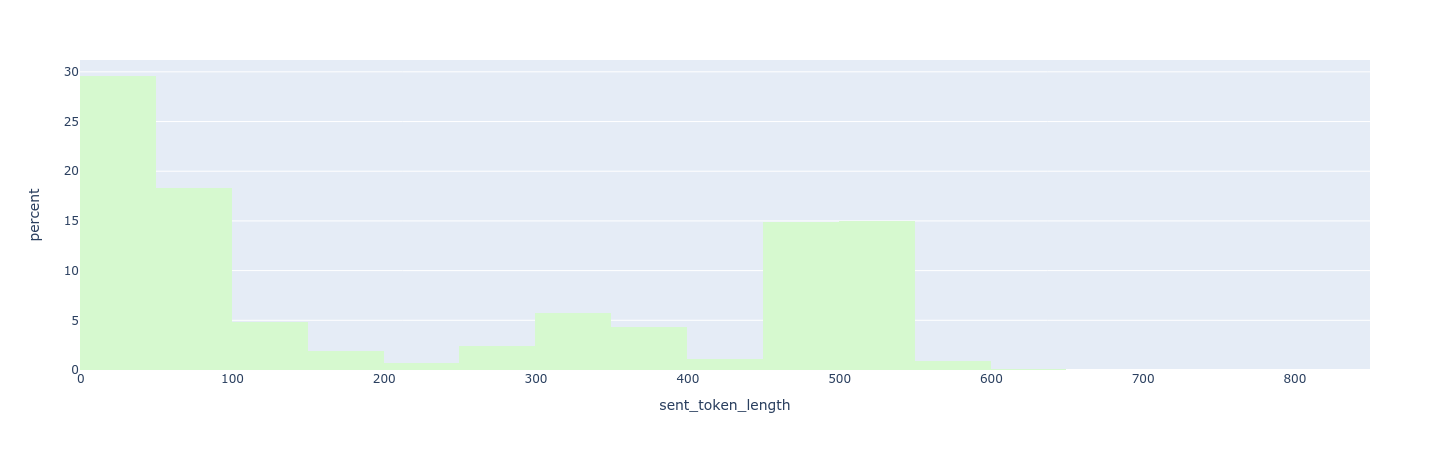}
     \caption{Number of tokens per sequence on B\&PEML dataset.}   
    \end{subfigure}
    \caption{Number of tokens per sequence on both datasets.}
    \label{fig:len_token}
\end{figure*}

According to \cite{trizna2023nebula}, it could be more suitable to use a smaller embedding size and increase the number of heads in the MHA layer. So we set the embedding size to 32, and we set the number of heads for each MHA layer to 8. We also choose to restrict the size of the input sequences given to the model to 512 tokens. This choice is justified by our analysis of the sequences length in both our datasets, presented in Figure \ref{fig:len_token}, which shows that very few sequences contain more than 500 tokens (only 12\% in the B\&PEML dataset).

\section{Results}
\label{sec:results}

\subsection{Tests on the different datasets}

We first train our model on Quo Vadis dataset only with a train-test separation as defined by their authors. Note that the testset provided by Quo Vadis is composed of 3 months-old data compared to its training data, as explained in \cite{trizna2022quo}. We retain 80\% of the train dataset for training and 20\% for validation. Our model achieves 98\% (resp. 97\%) accuracy on the train (resp. validation) dataset. Table \ref{tab:testQuo} summarizes the F1-score and the accuracy result computed on the test corpus of datasets detailed in Section \ref{subsec:dataset}. 



\begin{table}[!ht]
    \centering
    \caption{Results on different datasets using our model trained on Quo Vadis dataset only.}
    \label{tab:testQuo}
    \begin{tabular}{|c|c|c|}
        \hline
         & F1-score & Accuracy \\ 
         \hline
       Quo Vadis  & 0.87 & 0.872 \\
       \hline
       B\&PEML & 0.76 & 0.73 \\
       \hline
       OBMID-24 & 0.385 & 0.76 \\
       \hline
    \end{tabular}

\end{table}

\begin{figure}[!ht]
     \centering
     \begin{subfigure}{0.3\textwidth}
         \includegraphics[width=\textwidth]{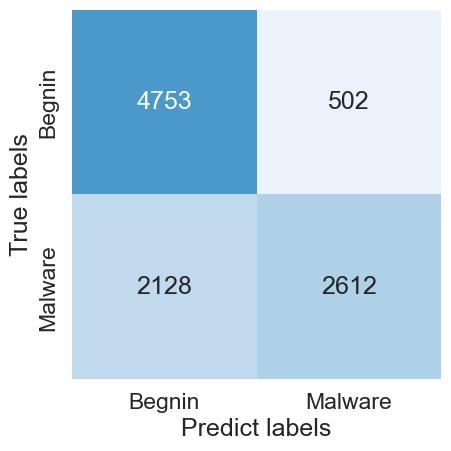}
     \caption{Test on the B\&PEML dataset.}
     \end{subfigure}
     \hspace{1cm}
     \begin{subfigure}{0.3\textwidth}
         \includegraphics[width=\textwidth]{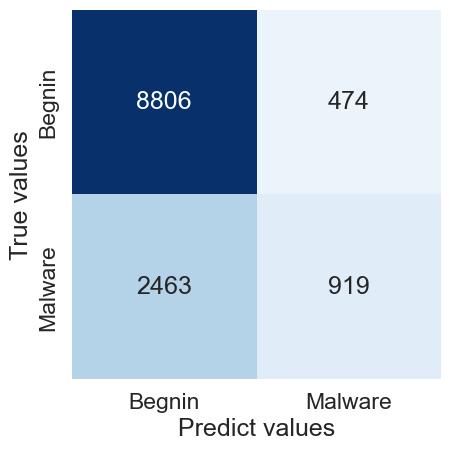}
     \caption{Test on the OBMID-24 dataset.}
     \end{subfigure}
         \caption{Confusion matrices for training on the Quo Vadis dataset.}
    \label{fig:conf_matrices_Quo}
 \end{figure}

 Results are pretty decent on the Quo Vadis dataset but quite disappointing on the two other datasets. On OBMID-24, our model performs poorly which can be interpreted as a low generalization capability or a possible overfitting problem. Figure \ref{fig:conf_matrices_Quo} shows the confusion matrices for these tests. There are also a lot of malware files tagged as benign files, which is not acceptable in operational conditions.

In the next experiment, we train our model on B\&PEML dataset with train-validation-test split of 70\%-15\%-15\%, stratified sampling, and early stopping. This model gives train and validation accuracies of 99\%. 

\begin{table}
    \centering
    \caption{Results on different datasets using our model trained on B\&PEML dataset only.}
    \label{tab:testBod}
    \begin{tabular}{|c|c|c|}
        \hline
         & F1-score & Accuracy \\ 
         \hline
       Quo Vadis  & 0.70 & 0.70 \\
       \hline
       B\&PEML & 0.99 & 0.99 \\
       \hline
       OBMID-24 & 0.63 & 0.85 \\
       \hline
    \end{tabular}
\end{table}

\begin{figure}[!ht]
    \centering
    \begin{subfigure}{0.3\textwidth}
        \includegraphics[width=\textwidth]{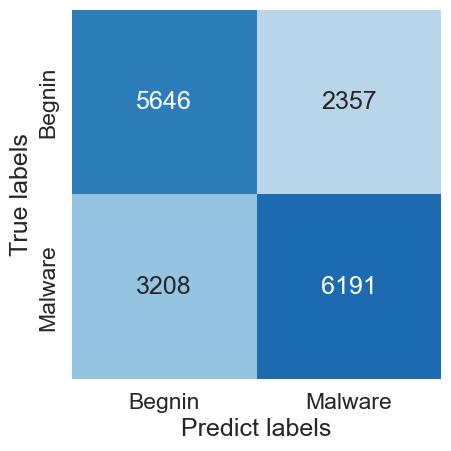}
        \caption{Test on the Quo Vadis dataset.}
    \end{subfigure}
        \begin{subfigure}{0.3\textwidth}
        \includegraphics[width=\textwidth]{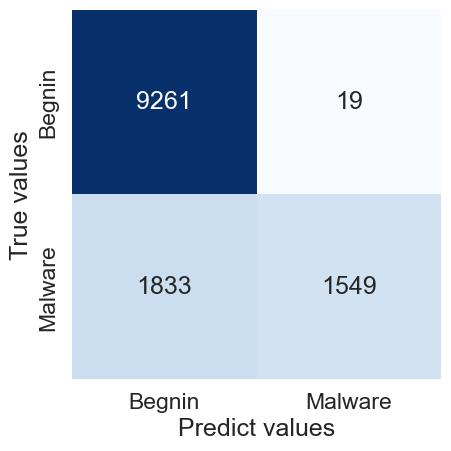}
            \caption{Test on the OBMID-24 dataset.}
    \end{subfigure}
   \caption{Confusion matrices for training on the B\&PEML dataset.}
    \label{fig:conf_matrices_Bod}
\end{figure}

Results on B\&PEML dataset are very good, but both F1-score and accuracy decrease significantly when inference is conducted on other datasets, which suggests a potential overfitting problem as in the first experiment. 
Figure \ref{fig:conf_matrices_Bod} shows the confusion matrices for this second experiment.

Our last test consists in training our algorithm on both B\&PEML and Quo Vadis datasets. After observing no significant changes in results, we choose to also use OBMID-24 as the validation subset. We get a train accuracy of 98.5\%, and a validation accuracy of 95\%. We test this model on B\&PEML and on QuoVadis, with the same test samples as for the two previous experiments. Results are summarized in Table \ref{tab:testBodQuo}. Note that we sidestep the overfitting problem by using validation data comprising very recent files, that were not processed by the model during training.

\begin{table}
    \centering
    \caption{Results with our model trained on the Quo Vadis and B\&PEML datasets and validated on OBMID-24 datasets.}
    \label{tab:testBodQuo}
    \begin{tabular}{|c|c|c|}
        \hline
         & F1-score & Accuracy \\ 
         \hline
       Quo Vadis  & 0.88 & 0.88 \\
       \hline
       B\&PEML & 0.99 & 0.99 \\
       \hline
    \end{tabular}
    
\end{table}

\begin{figure}[!ht]
    \centering
    \includegraphics[width=0.3\textwidth]{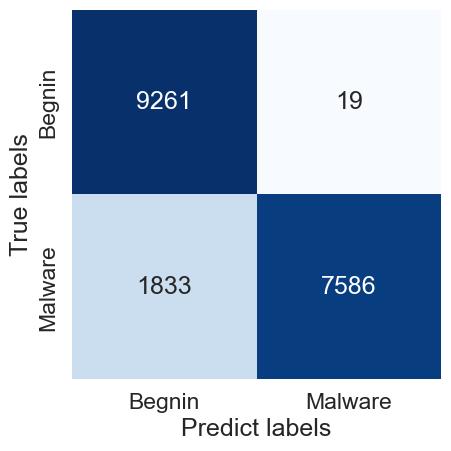}
    \caption{Confusion matrix for training on the Quo Vadis and B\&PEML datasets, validation on the OBMID-24 dataset and testing on Quo Vadis dataset.}
    \label{fig:trainBodQuotestQuo}
\end{figure}

This training increases the performance of our model for both datasets, even if there is still a significant amount of false negatives according to Figure \ref{fig:trainBodQuotestQuo} when testing the model on Quo Vadis. We also tested this training on both datasets without the validation step on recent files, which did not give us significant improvements.  

\subsection{Data analysis}
\label{subsec:data analysis}

\begin{figure*}[htb]
    \centering
    \begin{subfigure}[!ht]{\textwidth}
        \includegraphics[width=\textwidth]{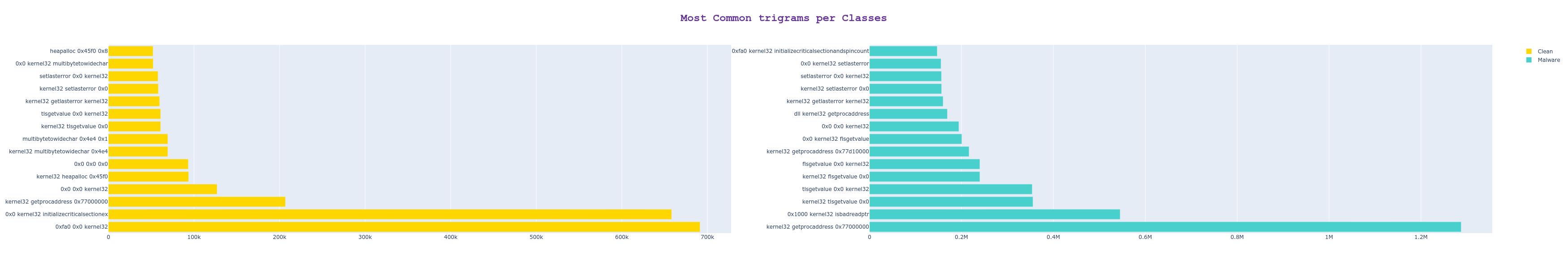}
        \caption{Trigram occurrences in Quo Vadis.}  
        \label{fig:trigram_Quo}
    \end{subfigure}
    
    \begin{subfigure}[!ht]{\textwidth}
        \includegraphics[width=\textwidth]{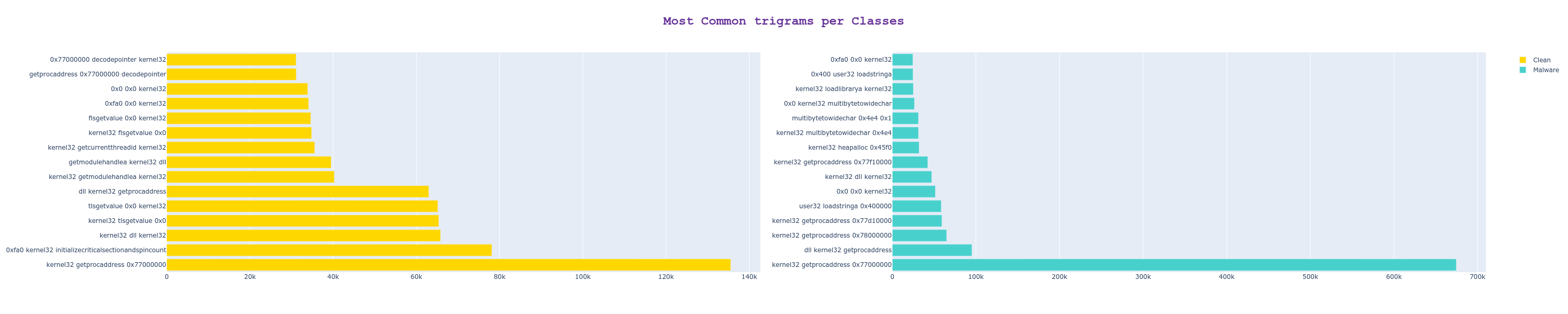}
        \caption{Trigram occurrences in B\&PEML.} 
        \label{fig:trigram_Bod}
    \end{subfigure}
    \caption{Trigram occurrences for both dataset on the train samples.}
    \label{fig:trigram}
\end{figure*}

In order to have a better understanding of our results, we conducted a deeper analysis of the different datasets used for our experiments. We compute the most recurrent trigrams per classes for both Quo Vadis and B\&PEML datasets, to see if the distribution is the same or if some differences between classes could lead to some bias or to some relevant information. Figure \ref{fig:trigram} presents the trigram occurrences in both datasets.

We can observe on Figure \ref{fig:trigram_Quo} that the trigram \textit{kernel32 getprocaddress 0x7000000} is the most common trigram in the malware samples of the Quo Vadis train set, while it is only the third common trigram in the benign samples, with a third of the occurrences of the first trigram in this case. As a consequence, this trigram may be interpreted by our model as the main information related to the maliciousness of a file. In Figure \ref{fig:trigram_Bod}, we see that this API call is the most common in both malware and benign samples on B\&PEML dataset. We suspect that this specific trigram might induce a bias when training our model on the Quo Vadis dataset, because the benign distribution of trigrams is not the same as its malware counterpart, for common API instructions. This potential bias could explain the aforementioned bad generalization capability when training our model on Quo Vadis and testing it on other datasets.

\subsection{Complementary analysis}
\label{subsec:complementary_analysis}

\begin{figure}[!ht]
     \centering
     \begin{subfigure}{0.5\textwidth}
         \centering
         \includegraphics[width=1\textwidth]{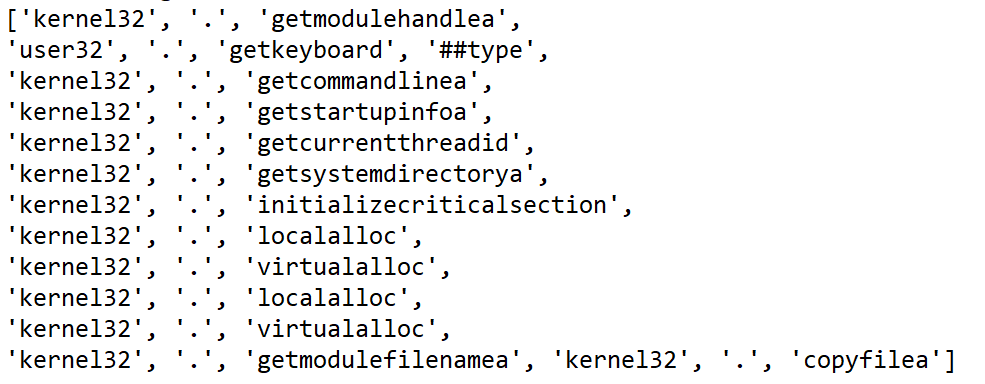}
         \vspace{0.05cm}
         \caption{Data format after preprocessing without arguments in the sequences of API calls.}
         \label{fig:preco_without_args}
     \end{subfigure}
     \hfill     \vspace{0.05cm}
     \begin{subfigure}{0.5\textwidth}
         \centering
         \includegraphics[width=1\textwidth]{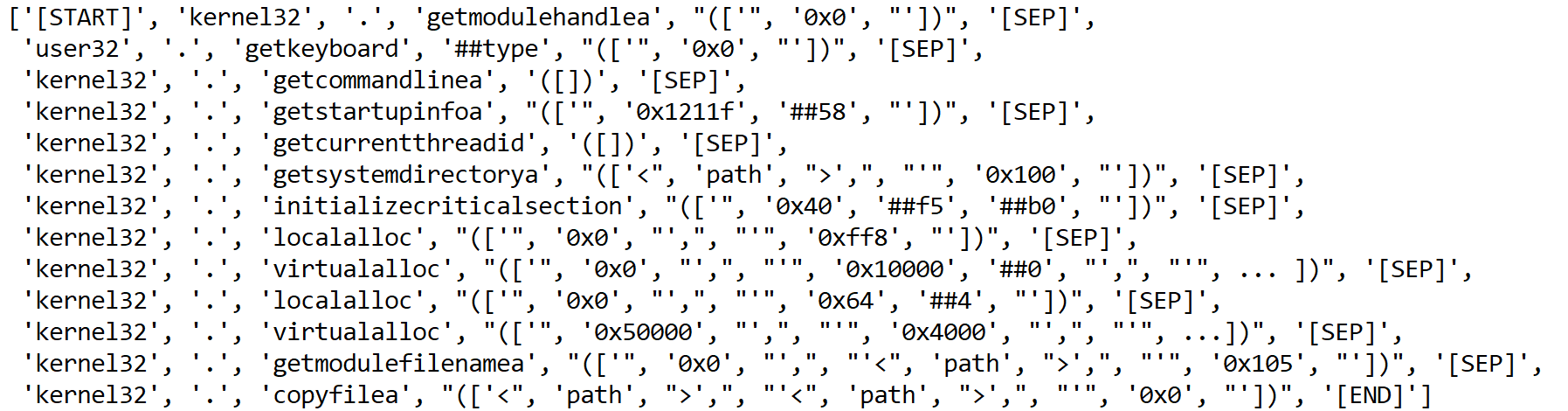}
         \vspace{0.05cm}
         \caption{Data format after adding tags as [START], [END] and [SEP].}
         \label{fig:preco_sep}
     \end{subfigure}
        \caption{Two other preprocessing methods used in our experimentation.}
        \label{fig:preco_2}
\end{figure}

We ran other tests on our model in order to challenge our findings. We trained our model with the API calls without the arguments, as shown in Figure \ref{fig:preco_without_args}. When training on Quo Vadis, the results of \cite{trizna2023nebula} are confirmed with a $9\%$ reduction in F1-score for the test on Quo Vadis, but on B\&PEML F1-score drops to $20\%$.  We also notice that the majority of binaries were classified as benign which is a major drawback. When training on B\&PEML, the conclusions are fairly similar. The B\&PEML test loses a few percentages but remains at $97\%$ in terms of F1-score and accuracy. Results on Quo Vadis dataset are somehow similar but on the OBMID-24 dataset, results drop to $72\%$ in F1-score and $53\%$ in accuracy. These results suggest that API calls arguments have a strong contribution and are therefore needed to generalize on data that differ from the training dataset. 

We have also experimented a third preprocessing method based on the sequence of API calls, as shown in Figure \ref{fig:preco_sep}, adding tags such as [START] and [END] or [SEP] to separate API calls and arguments from each other.  However, both modifications did not demonstrate any improvement from a performance standpoint. Worse, by adding the [SEP] tag every seven tokens on average and maintaining a length of 512 tokens, we are losing approximately ten instructions per report. 

We ran Nebula code available on Github\footnote{Nebula Github: \url{https://github.com/dtrizna/nebula}, July 2024}, directly on the B\&PEML dataset. The results are rather disappointing with a $40\%$ F1-score and an accuracy of $57\%$, suggesting a possible generalization problem. It would be interesting to dig deeper in future works to gain a better understanding of these results. 

\section{Limitations}
\label{sec:limitations}

Speakeasy is a valuable security solution since (i) it is capable of generating sandbox-like reports quickly and easily with a lot of quality information, (ii) it is an open-source project with a permissive software license (MIT license) and (iii) it is still active at the time of writing (July 2024). Nevertheless, it suffers a few limitations. As mentioned by their authors, it cannot fully emulate some files \cite{Speakyeasy-limitations} since windows API are too many to be all emulated. A given emulation may also fail due to a missing expected environment by the sample under analysis. This comprises for instance the absence of specific files or registry keys and also unexpected data as returned by the emulation engine. 

Our experiments faced these limitations with an impact on our computed model. Indeed, even if very few sequences from Speakeasy reports are longer than 500 tokens, we can see on Figure \ref{fig:len_token} that on both B\&PEML and Quo Vadis datasets lots of sequences have no more than 50 tokens: 18\% on Quo Vadis and 30\% on B\&PEML. Another meaningful metric is that 53\% of sequences exhibit less than 150 tokens, due to the fact that the emulation tool stops when it encounters an API call that is not emulated. This leads to low detailed reports that are less valuable when training our model. 

From an attacker perspective, this limitation is useful to break any Speakeasy analysis. An attacker would just have to call an unsupported API at the beginning of its malicious payload to bypass a detection algorithm, which would then analyze an insufficient detailed emulation report. Mitigation is still possible since emulated API handlers can be added by defining a Python function with the correct name and arguments even if this kind of mitigation is an unbalanced cat-and-mouse game. As a consequence, a detection model based on Speakeasy reports must not be used as the only line of defense in the context of PE files dynamic analysis.


\section{Discussion and Future Works}
\label{sec:future work}


Our approach using an SLM for malware detection allows us to use AI in a responsible way, and forces us to get a better knowledge about our data in order to implement more precise preprocessings and identify valuable features.  

According to our experiments, our model can really take advantage of being trained on a combination of datasets, while being challenged in its validation steps with more recent samples. We still need to conduct more investigations to improve our false-negatives ratio. We think this could be done for instance by cleaning and normalizing input data even more during the preprocessing step. 

Some additional data analysis are also required to make sure that the distributions of API calls of our malware and benign samples are similar, in order to prevent any bias in the data. 


The next step in our work consists in combining static and behavioral analysis into an ``hybrid" model. 
It would first use a machine learning model based on static features to identify the maliciousness of a sample \cite{marais:hal-03881198} then further analyze it using the Encoder-Only model in case this sample has not been classified with sufficient confidence. 
We expect this hybrid model to be more accurate than any model that would use either static or behavioral features. This hybrid model would also have the advantage of being rather small, which matches with our low carbon footprint side objective. 

Even if Speakeasy suffers some limitations (cf. Section \ref{sec:limitations}), it still remains a very interesting tool that should benefit from being complemented by other dynamic analysis techniques. Combining and correlating security information from such other techniques is a promising research field for some future works. 

We are also curious to assess the robustness of this Encoder-based model and of our future hybrid detection model when challenged with offensive tools such as Gym-Malware \cite{Anderson2017}, Malfox \cite{Zhong2020}, or our own adversarial model MERLIN \cite{quertier2022merlinmalwareevasion}. These tools are capable of evading some static antivirus detection engines by altering some key characteristics of a given PE file. It would be interesting to evaluate the impacts of such alterations on both Speakeasy emulation reports and on our detection results.

\section*{Conclusion}

In this paper, we designed a Small Language Model leveraging an Encoder-only architecture, for malware detection based on behavior analysis. We deliberately limited the size of our model for carbon footprint considerations and ease of deployment in an operational context. These restrictions allowed us to identify several key factors for this task, such as the importance of an optimal tokenizer, the efficiency of the emulation tool that reports dynamic behavior and as usual, data quality. Our results are pretty decent but suggest that some improvements should be conducted to enhance the performance of our detection chain. Besides, comparing our results on several datasets seems to indicate a bias for the Quo Vadis dataset, which could explain some bad generalization capabilities of our model. Finally, a good practice to avoid overfitting and maintain good generalization capabilities could be to rely on very recent samples during the validation step. 

\section*{Acknowledgments}

This work is supported by Orange FSI funding program coordinated by Adrien Servouze and Jean-Marc Blanco and also Orange Innovation SECMA research project led by Sok-Yen Loui. We would like to thank them for their valuable support. 

\clearpage

\bibliographystyle{unsrt}
\bibliography{ref}

\end{document}